\documentclass[aps,prx,reprint,showpacs,superscriptaddress,floatfix]{revtex4-2}

\usepackage{notoccite}
\usepackage{natbib}
\usepackage{graphicx}
\usepackage{dcolumn}
\usepackage{bm}
\usepackage{color,graphicx,subfigure}
\usepackage{enumitem}
\usepackage{wrapfig}
\usepackage{soul}
\usepackage{amsmath}
\usepackage{amssymb}
\usepackage{threeparttable}
\usepackage{siunitx}
\usepackage{multirow}
\usepackage{makecell}
\usepackage{booktabs}
\usepackage{textcomp}
\usepackage{gensymb}
\usepackage[]{hyperref}
\hypersetup{colorlinks=true,linkcolor=blue,citecolor=blue,urlcolor=blue,pdfpagemode=UseNone}
\usepackage{xcolor}
\usepackage{etoolbox}
\usepackage{floatrow}
\usepackage{floatflt}
\usepackage{datetime}
\usepackage{graphicx}
\usepackage{dcolumn}
\usepackage{bm}
\usepackage{lineno}
\usepackage[normalem]{ulem}

\begin{document}

\title{An altermagnetic materials library in intercalated transition-metal dichalcogenides}

\author{Ezra Day-Roberts}
\affiliation{\footnotesize Department of Physics, Arizona State University, Tempe, Arizona 85287, USA}

\author{Huan Wu}
\affiliation{\footnotesize Department of Physics, Arizona State University, Tempe, Arizona 85287, USA}

\author{Onur Erten}
\affiliation{\footnotesize Department of Physics, Arizona State University, Tempe, Arizona 85287, USA}

\author{A.\,S\, Botana}
\affiliation{\footnotesize Department of Physics, Arizona State University, Tempe, Arizona 85287, USA}

\begin{abstract}

Altermagnets represent a promising class of magnetic materials owing to their distinctive spin-split band structures in the absence of net magnetization. Here, we present a first-principles investigation of altermagnetism in magnetically intercalated transition metal dichalcogenides (TMDs) with the general formula T$_y$MX$_2$ (T= 3$d$-transition metal, M= transition-metal, X=chalcogen, $y$= 1/3 or 1/4). For a TMD host with 2H structure, compounds exhibiting A-type antiferromagnetism  are $g$-wave altermagnets by symmetry. We identify several intercalated TMDs fulfilling the conditions for altermagnetic order to be realized. Several of these candidate materials display spin-splittings at the Fermi level as large as 100 meV. 

\end{abstract}

\maketitle

\section{Introduction}
Altermagnetism (AM) represents an emerging magnetic phase that combines traits of ferromagnets and antiferromagnets: with time reversal symmetry-breaking and alternating spin-split band structures, alongside antiparallel magnetic order, and a vanishing net magnetization constrained by symmetry \cite{vsmejkal2022beyond, vsmejkal2022emerging}. As such, altermagnets can offer significant advantages for spintronic applications with their eV-scale spin splitting comparable to that of ferromagnets, combined with the THz spin dynamics and vanishing stray fields of antiferromagnets \cite{bai2024altermagnetism, wei2024crystal, mazin2024altermagnetism}.  As exciting as the prospects of altermagnetism are --and despite substantial theoretical progress-- the symmetry requirements to realize AMs are quite strict, and, as a consequence, just a handful of altermagnetic materials have been experimentally confirmed including MnTe \cite{Amin2024}, CrSb \cite{crsb}, Mn$_5$Si$_3$ \cite{mn5si3}, KV$_2$Se$_2$O \cite{KV2Se2O}, and $\alpha-$Fe$_2$O$_3$ \cite{Fe2O3}.

\begin{figure}
    \centering
    \includegraphics[width=\linewidth]{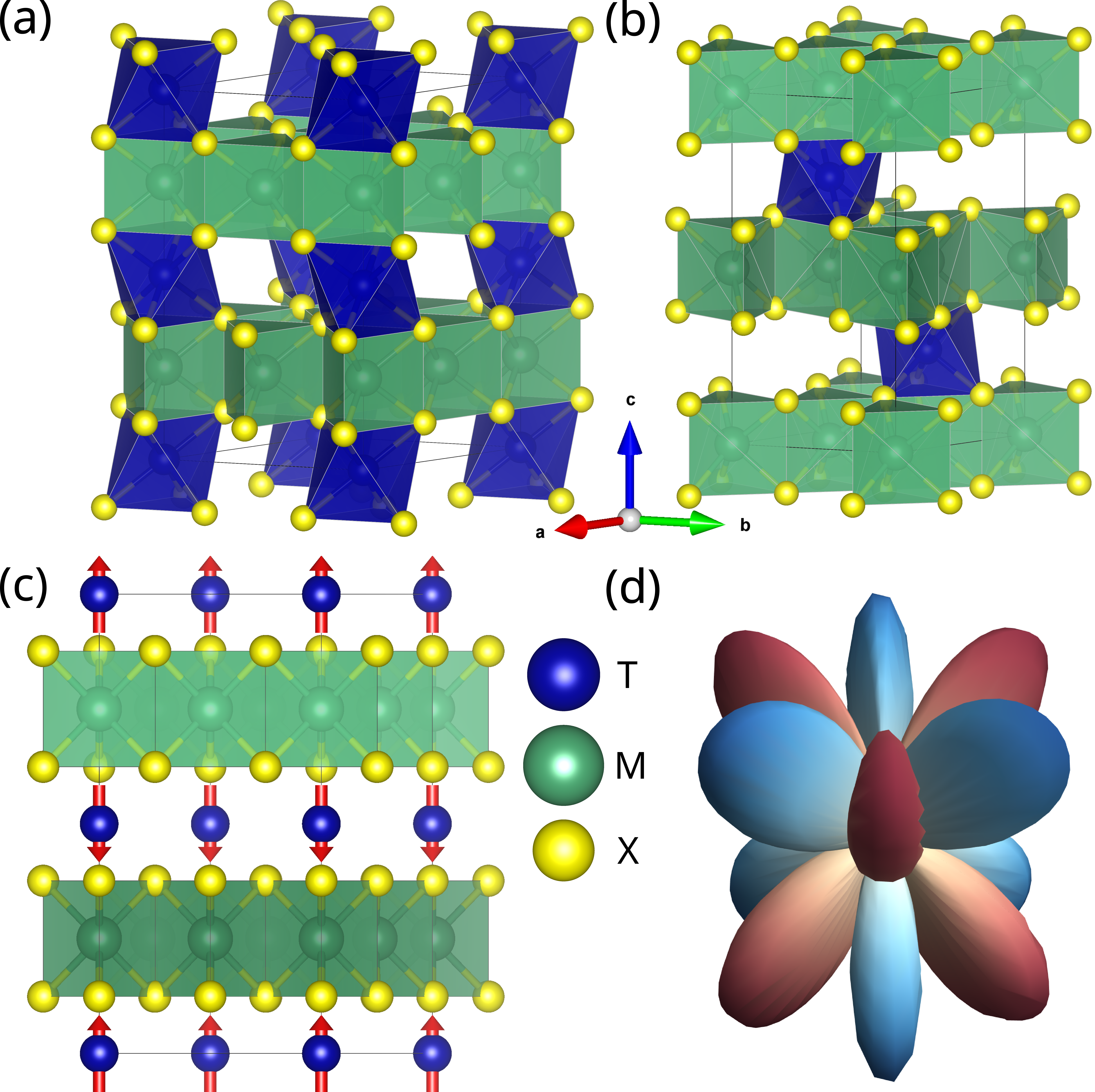}
    \caption{Crystal structure of (a) $y=1/4$ and (b) $y=1/3$ intercalated TMDs T$_y$MX$_2$ showing AA and AB intercalant stacking, respectively. (c) Antiferromagnetic AF-A type magnetic structure shown on the intercalant sublattice consisting of ferromagnetic planes coupled antiferromagnetically out of plane. AF-A order is displayed for $y=1/4$ filling but the $y=1/3$ case is analogous. AF-A order in the intercalant sublattice gives rise to altermagnetism in both cases. (d) Structure of $g$-wave altermagnetic splitting realized in intercalated TMDs.}
    \label{fig:schematic}
\end{figure}

 In the arena of 2D van der Waals (vdW) materials, intercalated transition-metal dichalcogenides (TMDs) MX$_2$ (M=transition metal, X=chalcogen) are particularly promising \cite{review_tmds}. The intercalation of 3$d$-transition metals (T) in MX$_2$ gives rise to a class of materials T$_y$MX$_2$ that are known to display a variety of magnetic phases, including easy axis/plane ferromagnetism and antiferromagnetism, or helical magnetic states \cite{cr_tas2, fe_nbs2, fe_nbs2_14, nbs2_cr_14, ni_nbs2, v_nbs2, PhysRevB.103.174431, PhysRevMaterials.4.054416, Ray2025, cr_nbs2, TogawaChiral2012,  KousakaChiral2009, VANLAAR1971154, KousakaChiral2009, mn_nbs2_nbse2_14, RahmanRKKY2022, PhysRevX.12.021003, Little2020, Gorochov01041981, TSUJI2001213, Lawrence2023, co_nbs2, PhysRevB.105.155114, Ghimire2018, Takagi2023, Battaglia2007, PhysRevB.108.054418, GubkinCrystal2016, VOORHOEVEVANDENBERG1971167, Parkin01011980, nbse2_cr_14, v_tas2, ObeysekeraMagneto2021, YamasakiExfoliation2017, mn_tas2, fe_tas2, fe_tas2_14, CheckelskyAnomalous2008, KoRKKY2011, Park2023, tas2_parkin, PhysRevB.109.085135, sprague2025}. Their magnetic behavior is strongly dependent on the superlattice order but also on the host matrix. In particular, the intercalation of magnetic atoms in 2H-TaS$_2$ and NbS$_2$/Se$_2$ has been more intensively scrutinized, with the most studied intercalant concentrations corresponding to 1/4 and 1/3 of the octahedral holes being occupied in an ordered manner \cite{cr_tas2, fe_nbs2, fe_nbs2_14, nbs2_cr_14, ni_nbs2, v_nbs2, PhysRevB.103.174431, PhysRevMaterials.4.054416, Ray2025, cr_nbs2, TogawaChiral2012,  KousakaChiral2009, VANLAAR1971154, KousakaChiral2009, mn_nbs2_nbse2_14, RahmanRKKY2022, PhysRevX.12.021003, Little2020, Gorochov01041981, TSUJI2001213, Lawrence2023, co_nbs2, PhysRevB.105.155114, Ghimire2018, Takagi2023, Battaglia2007, PhysRevB.108.054418, GubkinCrystal2016, VOORHOEVEVANDENBERG1971167, Parkin01011980, nbse2_cr_14, v_tas2, ObeysekeraMagneto2021, YamasakiExfoliation2017, mn_tas2, fe_tas2, fe_tas2_14, CheckelskyAnomalous2008, KoRKKY2011, Park2023, tas2_parkin, PhysRevB.109.085135, sprague2025}.
When the 3$d$ transition metal element occupies $y$ =1/4 of the octahedral voids, it forms a centrosymmetric structure with a 2 $\times$ 2 superlattice in the hexagonal space group P6$_3$/mmc. In contrast, when the intercalant atom occupies $y$=1/3 of the holes, it creates an ordered $\sqrt{3}$ $\times$ $\sqrt{3}$ superstructure within the non-centrosymmetric hexagonal spacegroup P6$_{3}$22, as depicted in Fig. \ref{fig:schematic}(a,b).

While intriguing transport
phenomena (such as the anomalous Hall effect) have been investigated in the family for years \cite{Ghimire2018, Tenasini2020_GiantAHE_Co13NbS2, Park2022_Co13TaS2_ToroidalAHE}, altermagnetism has only been recently discussed, particularly in the Co-intercalated NbSe$_2$ and TaSe$_2$-based compounds at $y$=1/4 \cite{Regmi2025_nbse214, bai2024altermagnetism, sprague2025}. In spite of some latest efforts \cite{Sah2025Altermagnetism, Tenzin2025Persistent}, the governing principles underlying the host- and intercalant-dependent magnetic properties in intercalated TMDs --and consequently the realization of altermagnetic spin splittings-- remain largely unknown. 

Here, we perform a comprehensive \textit{first-principles} characterization of T$_y$MX$_2$ compounds, where T = V, Cr, Mn, Fe, Co, Ni; $y$= 1/4, 1/3, M = Nb, Ta, and X = Se, S. Via a symmetry analysis, we show how an A-type antiferromagnetic (AF) order of the T sublattice satisfies the criteria for altermagnetism for both $y$= 1/3, 1/4 intercalant ratios for a 2H-TMD host and for $y$=1/3 for a 1T host. In both cases, the symmetry operation associated with connecting the two magnetic sublattices is a mirror or screw, but not an inversion. Our analysis from DFT calculations on 2H-intercalated TMDs subsequently identifies promising T$_y$MX$_2$ altermagnets with spin-splittings at the Fermi level as large as 100 meV.

\section{Computational methods}
The density functional theory (DFT) calculations were performed using the Vienna $Ab$ $initio$ Simulation Package (VASP) \cite{VASP} with the Perdew-Burke-Ernzerhof exchange-correlation functional and the projector-augmented wave method \cite{VASP-PAW,Perdew1996Generalized}. The energy cutoff for the
plane-wave basis set was 500 eV. A 12 $\times$ 12 $\times$ 6 $\Gamma$-centered $k$-point grid was used for the primitive cell.

We performed structural optimizations for T$_y$MX$_2$ compounds with a 2H host (T = V, Cr, Mn, Fe, Co, Ni; $y$= 1/4, 1/3, M = Nb, Ta, and X = Se, S). The van der Waals corrections were incorporated using the DFT-D3 method with the Becke-Johnson damping function \cite{DFT-D3}. The DFT+$U$ approach, following Dudarev's formulation \cite{Dudarev}, was applied with a Hubbard $U$ parameter of 3 eV (a $U$ was used as in some materials no net magnetic moments are obtained without a $U$). Cell parameters and atomic positions were optimized until all interatomic forces were less than $10^{-4}$ eV/\AA.  In this optimization, we assumed the spin configuration as ferromagnetic (FM), and the lattice parameters
and internal coordinates were optimized, keeping the original
space group symmetries P6$_3$22 for $y$ = 1/3 and P63/mmc
for $y$ = 1/4.  We subsequently performed DFT-based calculations to compare the energetics of a ferromagnetic (FM), A-type AF and stripe-AF state (AF-\textbf{Q}). The stripe-AF state is characteristic of several of the known 3$d$-intercalated TMDs \cite{TanakaAHE2022,WuTunable2022,ParkComposition2024}.

\section{Results}
\subsection{Structural details}

We present the crystal structures of intercalated 2H-TMDs in
Fig. \ref{fig:schematic}(a,b). There are two intercalated transition metals per unit
cell, which are located in the vdW gaps. The intercalated transition metals are surrounded by a distorted
octahedron formed by chalcogen atoms. The
intercalated transition metals form a hexagonal closed-packed
lattice when $y$ = 1/3, and a triangular lattice stacked along the
$c$-axis when $y$ = 1/4. The optimized cell parameters for all the structures we have calculated are summarized in Appendix \ref{app_a}.

\begin{table}[]
    \centering
    \begin{ruledtabular}
    \begin{tabular}{c|c|c|c|c|c|c}
        Stacking & $x$ & Space Group & G & Irrep & H & AM? \\\hline
        2H & 1/4 & P6$_3$/mmc (194) & 6/mmm & $B_{2g}$ & $\overline{3}$2/m & \checkmark\\
        2H & 1/3 & P6$_3$22 (182) & 622 & $B_1$ & 32 & \checkmark\\
        1T & 1/4 & P$\overline{3}$m1 (164) & & & \\
        1T & 1/3 & P$\overline{3}$1c (163) & $\overline{3}$m & $A_{1u}$ & 32 & \checkmark 
    \end{tabular}
    \end{ruledtabular}
    \caption{Symmetries for potential altermagnetism (AM) for intercalated TMDs T$_y$MX$_2$ in both the 1T and 2H polymorphs. $G$ and $H$ are the point group with and without the altermagnetic order parameter.}
    \label{tab:symmetry}
\end{table}

 The calculated lattice parameters exhibit excellent agreement with the experimental values, where available in the Inorganic Crystal Structure Database \cite{ICSD} and in the literature \cite{VANLAAR1971154, HULLIGER1970117, fe_tas2_14}. The maximum deviation for the in-plane lattice constant $a$ is 0.075 {\AA} for Fe-intercalated NbSe$_2$ at the intercalate concentration $y$=1/3, while the majority of discrepancies are below 0.05 \AA, well within the expected error margins for DFT-D3. The agreement for the calculated out-of-plane lattice constant $c$ is also satisfactory, with deviations below 0.37 \AA. Although the deviations in $c$ are larger than those in $a$, this is expected in layered materials of this nature due to the more subtle energy landscape of van der Waals-bonded layers. The strong agreement validates the chosen computational framework for modeling these intercalated compounds and for making predictions in those that have not yet been experimentally realized.

\begin{table*}
    \caption{Relative stability of competing magnetic configurations in intercalated TMDs T$_y$MX$_2$. 
    For each host (MX$_2$) and intercalant (T), we report the energy difference between an AF-A and a FM state and between the striped-AF-$\mathbf{Q}$ state ($\mathbf{Q} = (1/2,0,0)$) and a FM state (in meV per primitive cell) at $y$= 1/3 and 1/4.    
    The calculated magnetic ground state (Calc.) is compared with the experimental magnetic order (Exp.), where available.
    Missing experimental information is indicated by --. Uncertainty in the type of experimentally-observed AF state is indicated by AF-?. AF-other is used to denote antiferromagnetic states different from AF-A. All energies quoted are obtained at $U=0$ eV other than for the cases indicated with a * that are obtained at $U=$ 3 eV since some of the chosen magnetic states cannot be converged at $U$= 0 eV. The magnetic ground states with a $^\dagger$ change at $U$= 3 eV. }
    \begin{ruledtabular}
    \begin{tabular}{c|c|cccc|cccc}
        \multirow{2}{*}{Host} & \multirow{2}{*}{Int}
        & \multicolumn{4}{c|}{1/3 Intercalation}
        & \multicolumn{4}{c}{1/4 Intercalation} \\
        \cline{3-10}
        & &
        $E_{\mathrm{AF\text{-}A}} - E_{\mathrm{FM}}$ &
        $E_{\mathrm{AF\text{-}\mathbf{Q}}} - E_{\mathrm{FM}}$ &
        Calc. & Exp. &
        $E_{\mathrm{AF\text{-}A}} - E_{\mathrm{FM}}$ &
        $E_{\mathrm{AF\text{-}\mathbf{Q}}} - E_{\mathrm{FM}}$ &
        Calc. & Exp. \\
        \hline

        \multirow{6}{*}{NbS$_2$}
          & V  & -13.02340 &  -3.85560 & AF-A               & AF-A\cite{PhysRevB.103.174431, PhysRevMaterials.4.054416, Ray2025}     &  28.50130 &  25.95740 & FM$^\dagger$                 & --       \\
          & Cr &  81.70170 &  49.49910 & FM                 & HM\cite{cr_nbs2, TogawaChiral2012,  KousakaChiral2009}       & 168.75200 & 129.67300 & FM & AF-A\cite{VANLAAR1971154}     \\
          & Mn &  75.76870 &  23.97230 & FM                 & HM\cite{KousakaChiral2009}       &  25.29130 &  52.41400 & FM                 & FM\cite{mn_nbs2_nbse2_14, RahmanRKKY2022}        \\
          & Fe & -23.86140 & -60.53330 & AF-$\mathbf{Q}^\dagger$   & AF-other\cite{PhysRevX.12.021003, Little2020} & -113.76800 & -26.57830 & AF-A$^\dagger$           & AF-A\cite{Gorochov01041981, TSUJI2001213, Lawrence2023}     \\
          & Co & -20.72310 & -19.12323 & AF-A$^\dagger$                & AF-other\cite{co_nbs2, PhysRevB.105.155114, Ghimire2018, Takagi2023} & -82.89290 & -30.40960 & AF-A             & --       \\
          & Ni & -24.69560 &   6.82930 & AF-A$^\dagger$               & AF-other\cite{Battaglia2007, PhysRevB.108.054418} &   88.32693 & 56.62010  & FM*   & --       \\
        \hline

        \multirow{6}{*}{NbSe$_2$}
          & V  &  -0.70944 &  -0.06456 & AF-A             & --       &   8.09401 &   7.35305 & FM$^\dagger$                 & --       \\
          & Cr &  79.91460 &  48.44830 & FM                 & FM\cite{GubkinCrystal2016}        & 150.71200 & 111.27800 & FM                 & AF-?\cite{VOORHOEVEVANDENBERG1971167}    \\
          & Mn &  30.55597 &  -4.37023 & AF-$\mathbf{Q}$   & --       &  -7.94778 &   6.26362 & AF-A$^\dagger$              & HM\cite{mn_nbs2_nbse2_14}       \\
          & Fe & -51.94710 &  61.70090 & AF-A            & AF-?\cite{VOORHOEVEVANDENBERG1971167, Parkin01011980}    & -121.98000 & -39.17100 & AF-A               & AF-?\cite{nbse2_cr_14, Parkin01011980}    \\
          & Co & -37.14900 & -35.04087 & AF-A           & --       & -105.38800 & -48.37820 & AF-A               & --       \\
          & Ni & -32.25250 &  13.93780 & AF-A               & --       &   60.704630 &   41.505595 & FM*                 & --       \\
        \hline

        \multirow{6}{*}{TaS$_2$}
          & V  &  -9.67180 & -11.31310 & AF-$\mathbf{Q}$    & AF\cite{v_tas2}       &   4.21997 &   3.76207 & FM$^\dagger$                 & --       \\
          & Cr &  87.34290 &  47.46410 & FM                 & HM\cite{ObeysekeraMagneto2021, YamasakiExfoliation2017}       & 167.52600 & 125.74000 & FM                 & --       \\
          & Mn &  80.94110 &  31.38500 & FM                 & FM\cite{mn_tas2}        &  34.42740 &  55.82570 & FM                 & FM\cite{mn_nbs2_nbse2_14, Parkin01011980}        \\
          & Fe & 257.21900 & 115.97900 & FM               & FM\cite{fe_tas2}        &  -97.80455 & -23.31421 & AF-A$^\dagger$                & FM\cite{fe_tas2_14, CheckelskyAnomalous2008, KoRKKY2011}        \\
          & Co & -11.83326 & -13.76630 & AF-$\mathbf{Q}$    & AF-other\cite{co_nbs2, Park2023} & -94.29090 & -57.25040 & AF-A               & --       \\
          & Ni & -38.56840 &   7.88370 & AF-A               & AF-A\cite{tas2_parkin, PhysRevB.108.054418, PhysRevB.109.085135}     &  74.23060 &  49.83620 & FM                 & --       \\
        \hline

        \multirow{6}{*}{TaSe$_2$}
          & V  &   1.51296 &  -8.03755 & AF-$\mathbf{Q}$    & --       &   5.66487 &   0.64847 & FM$^\dagger$                 & --       \\
          & Cr &  89.27130 &  46.43700 & FM                 & --       & 148.27500 & 106.22600 & FM                 & --       \\
          & Mn &  36.44800 &   9.48596 & FM                 & --       & -14.54110 & 2349.20890 & AF-A$^\dagger$            & --       \\
          & Fe & -23.88390 &  -9.61780 & AF-A               & --       &  -46.72341 &  -129.23004 & AF-A*               & --       \\
          & Co &  -0.50630 & -14.29480 & AF-$\mathbf{Q}^\dagger$    & --       & -136.39500 & -89.51500 & AF-A               & --       \\
          & Ni & -46.01190 &   6.69400 & AF-A               & --       &  90.29764 &  69.62831 & FM*                & FM\cite{sprague2025}        \\

    \end{tabular}
    \end{ruledtabular}
    \label{tab:full}
\end{table*}

\subsection{Symmetry considerations for altermagnetism in intercalated TMDs} 
 While we focus our first-principles calculations on 2H-intercalated TMDs, we also lay the general symmetry considerations to realize altermagnetism in the 1T polymorph type. There are then four distinct structures, combining either a 1T or 2H TMD with a $y$= 1/4 or 1/3 intercalation fraction, as listed in Table \ref{tab:symmetry}. 
 A-type antiferromagnetism (shown in Fig.\ref{fig:schematic}(c)) gives rise to altermagnetic order if the two intercalant sites in a two layer cell are of the same Wyckoff position and related by a symmetry operation other than inversion or translation. For 2H TMDs at 1/4 and 1/3 intercalation, as well as for 1T with 1/3 intercalation, this criteria is met. It fails in the 1/4 intercalated 1T structure, because the quarter intercalation favors AA stacking and for a 1T host, it only has a single layer unit cell. Hence, the sites are related by a translation along $c$ and the magnetic ordering is $(0,0,\pi)$ rather than the $\Gamma$-point order needed for altermagnetism.  In the 1T structure at 1/3 intercalation, the intercalants order with AB stacking, generating a two-layer unit cell instead. The 2H structure already has a two-layer unit cell. In all three latter cases there is at least one symmetry relating the two intercalant sites. For 2H at $y$=1/4, they are connected by a six-fold screw axis along $(0,0,1)$, a two-fold screw axis along $(1,-1,0)$, as well as the inversion partner glides of these two operations. For 2H at $y$=1/3 there are the same two operations, but without the inversion partners. For 1T at $y$= 1/3, there is a single glide consisting of a mirror perpendicular to $(1,-1,0)$ combined with a translation of $(0,0,\frac{1}{2})$. In all of these cases, the altermagnetic pattern realized in the electronic band splitting, is 3-dimensional of $g$-wave kind, as shown in Fig. \ref{fig:schematic}(d).

\subsection{Magnetic states}

We summarize the energies of the DFT calculations for different magnetic states (FM, AF-A, AF-\textbf{Q}) together with the experimentally observed magnetic
structures for T$_y$MX$_2$ in Table \ref{tab:full}. As mentioned above, we analyze $y$= 1/3 and 1/4 intercalate levels in 2H TMDs NbS$_2$, NbSe$_2$, TaS$_2$, and TaSe$_2$; and V, Cr, Mn, Fe, Co, Ni are explored as 3$d$ intercalants. We quote the energies at $U$= 0 eV in Table \ref{tab:full}, but also specify in what systems a $U$ is needed to stabilize the chosen magnetic states, and in what cases a $U$= 3 eV changes the magnetic ground state. The experimental magnetic ground state (for the combinations that have been experimentally realized) is successfully reproduced in most of the DFT calculations, in agreement with previous work \cite{hatanaka2023prb}. 
It should be noted that we did not consider intralayer AF
states in the DFT calculations more complicated than a stripe with $\mathbf{Q}$= (1/2, 0, 0), neither did we consider the possibility of a helimagnetic state. In most of the systems with an experimentally confirmed helimagnetic ground state, the DFT calculations stabilize a FM state instead.  We ultimately identify the following as the
compounds that exhibit robust A-type AF
ground states (and are hence altermagnetic candidates) from the DFT calculations: V$_{1/3}$NbS$_2$, V$_{1/3}$NbSe$_2$, Fe$_{1/3}$NbSe$_2$, Co$_{1/3}$NbSe$_2$, Ni$_{1/3}$NbSe$_2$, Ni$_{1/3}$TaS$_2$, Fe$_{1/3}$TaSe$_2$, Ni$_{1/3}$TaSe$_2$, Fe$_{1/4}$NbS$_2$, Co$_{1/4}$NbS$_2$, Fe$_{1/4}$NbSe$_2$, Co$_{1/4}$NbSe$_2$, Co$_{1/4}$TaS$_2$, Fe$_{1/4}$TaSe$_2$, and Co$_{1/4}$TaSe$_2$. DFT calculations do not reproduce the observed experimental AF-order in Cr$_{1/4}$NbS$_2$ and Cr$_{1/4}$NbSe$_2$. In order to analyze the nature and trends of the magnetic ground states, we first discuss the valence states of the different intercalants. From the magnetic moments we obtain, V and Cr ions seem to be trivalent while the Mn, Fe, Co and Ni ions are divalent. Caution should be taken when trying to assign formal valence states in such itinerant systems but our findings seem in agreement with reported transport, optical and X-ray diffraction measurements \cite{Parkin01011980}.

The derived magnetic states reflect complex behavior that likely arises from the
competition between multiple exchange interactions. Given the metallic character of these
compounds, one might initially expect the dominant interaction to be the long-ranged
Ruderman--Kittel--Kasuya--Yosida (RKKY) exchange, whose oscillatory nature and sensitivity to both sign and magnitude could account for the diversity of observed magnetic orders.
However, even just by looking at the available experimental trends, RKKY exchange alone might be insufficient and it is likely that two magnetic interactions are at play: RKKY and superexchange via the orbitals on the chalcogen ions within the layers.

\begin{table}[]
    \centering
    \begin{ruledtabular}
    \begin{tabular}{c|c|c|c}
    material & at FL & $\pm$ 0.1 eV & $\pm$ 0.25 eV \\\hline
Fe$_{1/3}$NbSe$_2$   &  42   &  74   &  74   \\
Co$_{1/3}$NbSe$_2$   &  49   & 126   & 133   \\
Ni$_{1/3}$NbSe$_2$   &  58   &  63   & 114   \\\hline
Ni$_{1/3}$TaS$_2$    &  98   &  98   &  98   \\
Fe$_{1/3}$TaSe$_2$   &  17   &  56   &  81   \\
Ni$_{1/3}$TaSe$_2$   &  50   &  59   & 109   \\ \hline\hline

Cr$_{1/4}$NbS$_2$*        &    16 &    57 &    77\\
Fe$_{1/4}$NbS$_2$        &    65 &    92 &   152\\
Co$_{1/4}$NbS$_2$        &    67 &   102 &   140\\
Cr$_{1/4}$NbSe$_2$*       &   123 &   123 &   123\\
Fe$_{1/4}$NbSe$_2$       &    51 &    89 &   151\\
Co$_{1/4}$NbSe$_2$       &    86 &   119 &   119\\\hline
Fe$_{1/4}$TaSe$_2$       &    10 &    11 &    13\\
Co$_{1/4}$TaSe$_2$       &   114 &   118 &   164\\
    \end{tabular}
    \end{ruledtabular}
    \caption{Maximum spin-splittings across the Fermi level (FL) or within a given energy range (0.1 or 0.25 eV) of the Fermi level for intercalated TMDs (T$_y$MX$_2$) identified as altermagnetic. Only materials with a spin-splitting above 10 meV at the FL are shown. Cr$_{1/4}$NbSe$_2$ and Cr$_{1/4}$NbS$_2$ are marked with a * as they are not found to be AF-A at $U=0$ eV in the DFT calculations. For Cr$_{1/4}$NbSe$_2$ the maximum splitting is found at the FL. Note that these splittings occur at generic, low symmetry positions.}
    \label{tab:splittings}
\end{table}

For example, at $y$= 1/3, a clear trend for dominant AF-A magnetic ground states can be observed to the right of the $3d$ row (Fe, Co, Ni). Alongside, the strength of both $180^\circ$ and $90^\circ$ superexchange interactions increases
systematically across the first-row transition metals. As the $d$ orbitals
become progressively filled, the $t_{2g}$--$e_g$ interactions are reduced and the overall
superexchange interaction is enhanced. This behavior reflects the dominant role of the $e_g$ orbitals, which participate in strong $\sigma$-bonding with the chalcogen ions. 
 As such, in Ni$^{2+}$, the $t_{2g}$ orbitals are
fully occupied, and the only allowed superexchange pathways are between neighboring $e_g$
orbitals, leading to strong antiferromagnetic interactions.
In contrast, Mn$^{2+}$ has half-filled $t_{2g}$ orbitals, allowing superexchange between an
$e_g$ orbital on one ion and a $t_{2g}$ orbital on another. These interactions involve
weaker $\pi$-bonding and are opposite in sign to the stronger $e_g$--$e_g$ interactions,
resulting in substantial cancellation and a very small net superexchange. Consequently,
superexchange interactions increase
progressively from Mn through Fe and Co to Ni. 

\begin{figure}
   \centering
   \includegraphics[width=\linewidth]{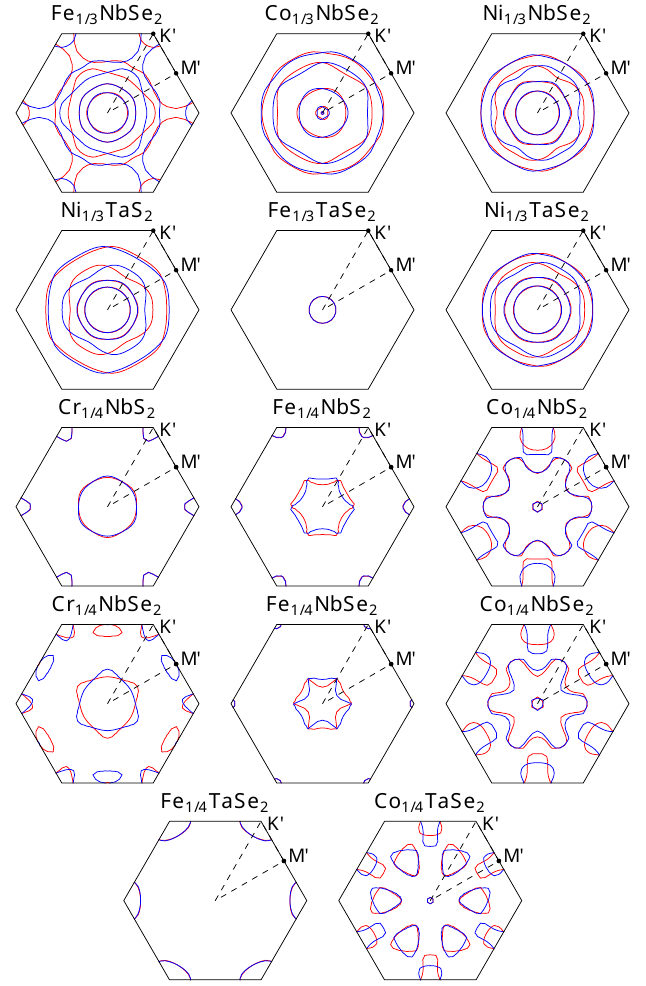}
   \caption{Fermi surface at $k_z=\pi/4c$ for the altermagnetic candidate materials found in intercalated TMDs T$_y$MX$_2$.}
    \label{fig:fermi_surfaces}
\end{figure}

In contrast, key aspects of the magnetic behavior of some intercalates are more
consistent with an RKKY-type interaction. The sign of the RKKY exchange depends
sinusoidally on $2k_F d$, where $d$ is the inter-ion spacing and $k_F$ is the Fermi
wavevector. While $d$ is similar across all intercalates, $k_F$ depends sensitively on the
filling of the $d$-derived conduction band, which in turn is controlled by the
concentration and oxidation state of the intercalant ion. For the Cr
and Mn $y=1/3$ intercalates, the $d$ sub-band is nearly full, resulting in a small $k_F$. Under
these conditions, the RKKY interaction is expected to favor ferromagnetic coupling, in
agreement with the experimentally (and DFT)-observed dominant ferromagnetic behavior of these
compounds.

\subsection{Spin splitting in altermagnetic candidates}

Intercalated TMDs with an A-type antiferromagnetic (AF) ground state for the intercalant ions, in which spins align
ferromagnetically within individual layers but antiferromagnetically between adjacent layers,
are altermagnetic, as explained above. This magnetic state leads to a compensated magnetic structure
with zero net magnetization per unit cell, yet the combination of magnetic ordering and
crystal symmetry permits a momentum-dependent spin splitting. The altermagnetic spin-split
bands are clearly visible in the band structure calculations shown in Appendix \ref{app_b} and in particular in the Fermi surfaces of Fig. \ref{fig:fermi_surfaces}. The 1/3 case has its nodal planes along $\Gamma$-M while the 1/4 case has them along $\Gamma$-K. The maximum spin-splittings at the Fermi level shown in Table \ref{tab:splittings} reveal several candidates with large spin-splittings at the Fermi level, the largest ones ($\geq$ 50 meV) obtained in Co$_{1/3}$NbSe$_2$, Ni$_{1/3}$NbSe$_2$, Ni$_{1/3}$TaS$_2$, Ni$_{1/3}$TaSe$_2$, Fe$_{1/4}$NbS$_2$, Co$_{1/4}$NbS$_2$, Cr$_{1/4}$NbSe$_2$, Fe$_{1/4}$NbSe$_2$, Co$_{1/4}$NbSe$_2$, and Co$_{1/4}$TaSe$_2$. Doping can offer further possibilities to increase the spin-splitting, as reflected in the maximum values that can be obtained within a 0.25 eV range around the Fermi level for several of the compounds we have calculated.

\section{Conclusions}

In conclusion, we have carried out a symmetry and \textit{first-principles} analysis of altermagnetism in magnetically intercalated transition-metal dichalcogenides with the general formula $T_yMX_2$. We have shown that for 2H TMD hosts, compounds with $y$=1/3 and 1/4 exhibiting A-type antiferromagnetism are symmetry-enforced $g$-wave altermagnets. Based on our electronic-structure calculations, we identify several candidate materials in which altermagnetic order can be realized: V$_{1/3}$NbS$_2$, V$_{1/3}$NbSe$_2$, Fe$_{1/3}$NbSe$_2$, Co$_{1/3}$NbSe$_2$, Ni$_{1/3}$NbSe$_2$, Ni$_{1/3}$TaS$_2$, Fe$_{1/3}$TaSe$_2$, Ni$_{1/3}$TaSe$_2$, Cr$_{1/4}$NbS$_2$, Fe$_{1/4}$NbS$_2$, Co$_{1/4}$NbS$_2$, Cr$_{1/4}$NbSe$_2$, Fe$_{1/4}$NbSe$_2$, Co$_{1/4}$NbSe$_2$, Co$_{1/4}$TaS$_2$, Fe$_{1/4}$TaSe$_2$, and Co$_{1/4}$TaSe$_2$. Several of these candidates exhibit spin splittings at the Fermi level reaching values of up to 100 meV (such as  Ni$_{1/3}$TaS$_2$, Co$_{1/4}$TaSe$_2$, and Cr$_{1/4}$NbSe$_2$) highlighting magnetically intercalated TMDs as a promising materials platform for realizing and exploiting altermagnetism in 2D vdW materials.

\begin{acknowledgements}
We acknowledge NSF Grant No. DMR-2206987 and the ASU Research Computing Center for HPC resources. 
\end{acknowledgements}

\bibliography{references}

\onecolumngrid

\appendix
\newpage

\section{Further structural details}\label{app_a}

Table \ref{tab:lattice_constants} presents the lattice constants for each of the intercalated TMDs analyzed in this work after DFT structural optimization in a FM state using the computational methodology described in the main text. 

\begin{table*}
    \caption{Lattice constants for intercalated TMDs. Experimental\textsuperscript{a} (Exp.) and calculated (Calc.) values are shown for both 1/3 and 1/4 intercalation concentrations. Missing experimental data are indicated by --.          $^a$All experimental lattice constants, unless otherwise stated, are sourced from the Inorganic Crystal Structure Database \cite{ICSD}. $^b$Taken from Ref.~\cite{VANLAAR1971154}.  $^c$Taken from Ref.~\cite{HULLIGER1970117}. $^d$Taken from Ref.~\cite{fe_tas2_14}.}
    \begin{ruledtabular}
        \begin{tabular}{c|c|cccc|cccc}                
        
        \multirow{2}{*}{Host} & \multirow{2}{*}{Intercalant} & \multicolumn{4}{c|}{1/3 Intercalation} & \multicolumn{4}{c}{1/4 Intercalation} \\
        \cline{3-10}
        & & \multicolumn{2}{c}{$a$ (\AA)} & \multicolumn{2}{c|}{$c$ (\AA)} & \multicolumn{2}{c}{$a$ (\AA)} & \multicolumn{2}{c}{$c$ (\AA)} \\
        \cline{3-10}
        & & Exp. & Calc. & Exp. & Calc. & Exp. & Calc. & Exp. & Calc. \\
        \hline        
        
        \multirow{6}{*}{NbS$_2$} 
          & V   & 5.7387\textsuperscript{b}   & 5.7278 & 12.1126\textsuperscript{b}   & 12.1473 & 6.641  & 6.5644 & 12.102  & 11.9263 \\
          & Cr  & 5.741  & 5.7100   & 12.097  & 12.0575 & 6.644   & 6.5936 & 11.914   & 11.9033 \\
          & Mn  & 5.7794\textsuperscript{b} & 5.7528 & 12.5993\textsuperscript{b} & 12.3992 & 6.68    & 6.6305 & 12.47   & 12.2355 \\
          & Fe  & 5.7609\textsuperscript{b} & 5.7526 & 12.1786\textsuperscript{b} & 11.9136 & 6.61   & 6.6268 & 12.08   & 11.8635 \\
          & Co  & 5.749  & 5.7330  & 11.886  & 11.8093 & 6.64   & 6.6052 & 11.848  & 11.7614 \\
          & Ni  & 5.7518\textsuperscript{b}  & 5.7095 & 11.8796\textsuperscript{b}  & 11.5556 & 6.64   & 6.5898 & 11.877  & 11.5068 \\
        \hline
        
        \multirow{6}{*}{NbSe$_2$}
          & V   & 5.99   & 5.9827 & 12.68   & 12.6922 & --     & 6.8881 & --      & 12.6563 \\
          & Cr  & 5.976  & 5.9582 & 12.567  & 12.6842 & 6.904  & 6.8760  & 12.57   & 12.5403 \\
          & Mn  & 6.008  & 6.0020  & 13.033  & 12.9587 & 6.942  & 6.9101 & 13.042  & 12.8249 \\
          & Fe  & 5.939  & 6.0135 & 12.671  & 12.4684 & 6.932  & 6.9031 & 12.702  & 12.4440  \\
          & Co  & 5.986  & 5.9808 & 12.384  & 12.3900   & 6.928  & 6.8847 & 12.431  & 12.3574 \\
          & Ni  & 5.986  & 5.9599 & 12.413  & 12.3288 & 6.911  & 6.8687 & 12.421  & 12.2922 \\
        \hline
        
        \multirow{6}{*}{TaS$_2$}
          & V   & 5.727  & 5.7151 & 12.201  & 12.2023 & --     & 6.5889 & --      & 12.1473 \\
          & Cr  & 5.72   & 5.6970  & 12.128  & 12.1253 & --     & 6.5800   & --      & 11.9789 \\
          & Mn  & 5.751\textsuperscript{c}  & 5.7404 & 12.508\textsuperscript{c}  & 12.4781 & 6.645\textsuperscript{b}  & 6.6215 & 12.552\textsuperscript{b}  & 12.3312 \\
          & Fe  & 5.7383\textsuperscript{b} & 5.7226 & 12.2392\textsuperscript{b} & 12.0831 & 6.6141\textsuperscript{d} & 6.6229 & 12.154\textsuperscript{d}  & 11.9770  \\
          & Co  & 5.7393\textsuperscript{b} & 5.7252 & 11.9329\textsuperscript{b} & 11.8661 & --     & 6.5986 & --      & 11.8345 \\
          & Ni  & 5.7336\textsuperscript{b} & 5.7114 & 11.9353\textsuperscript{b} & 11.7874 & --     & 6.5935 & --      & 11.7571 \\
        \hline
        
        \multirow{6}{*}{TaSe$_2$}
          & V   & 5.961\textsuperscript{c}   & 5.9664 & 12.743\textsuperscript{c}   & 12.7758 & --     & 6.8727 & --      & 12.7121 \\
          & Cr  & 5.954\textsuperscript{c}  & 5.9458 & 12.700\textsuperscript{c}    & 12.7810  & 6.870\textsuperscript{c}   & 6.8636 & 12.638\textsuperscript{c}  & 12.6412 \\
          & Mn  & --     & 5.9899 & --      & 13.0468 & --     & 6.9011 & --      & 12.9113 \\
          & Fe  & --     & 5.9196 & --      & 12.1427 & --     & 6.8282 & --      & 12.1661 \\
          & Co  & --     & 5.9008 & --      & 12.2296 & --     & 6.8776 & --      & 12.4432 \\
          & Ni  & --     & 5.9412 & --      & 12.2793 & --     & 6.8622 & --      & 12.3810  \\
        
        \end{tabular}        
    \end{ruledtabular}
    \label{tab:lattice_constants}
\end{table*}

\section{Bandstructures for candidate altermagnetic materials}\label{app_b}

\begin{figure}
    \centering
    \includegraphics[width=\linewidth]{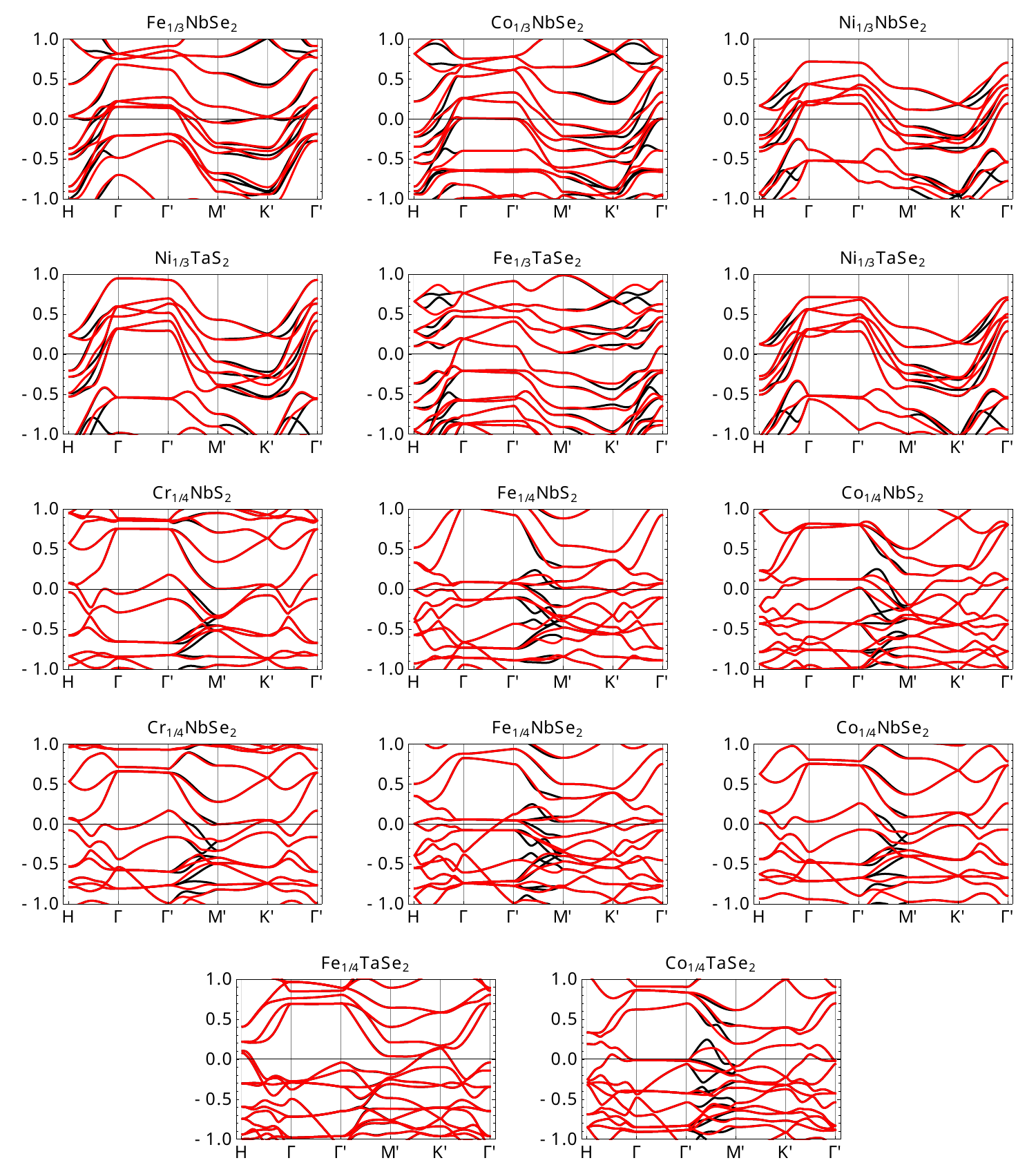}
    \caption{Band structures for candidate materials. Primed k-point labels correspond to coordinates in the $k_z=\pi/4c$ plane. Black and red lines are different spins.}
    \label{fig:bandstructures}
\end{figure}

Fig. \ref{fig:bandstructures} shows the band structures of all altermagnetic candidates in intercalated TMDs found in this work with sizable spin-splittings at the FL. 

\end{document}